\documentclass[nofootinbib,twocolumn,prc,aps]{revtex4-1}

\usepackage{enumerate}
\usepackage{amsmath}
\usepackage{amsfonts}
\usepackage{amsthm}
\usepackage{color}
\usepackage{graphicx}
\usepackage{array,multirow,graphicx}
\usepackage{hyperref}

\newcommand{\der}[2]{\frac{\partial #1}{\partial #2}}

\def\eps{\varepsilon}

\def\MeV{\rm ~MeV}    
\mathchardef\smin="2D

\begin{document}

\title{Anomalous quartic term in the expansion of the symmetry energy}

\author{Noemi Zabari}
\affiliation{Henryk Niewodnicza\'nski Institute of Nuclear Physics,\\
Polish Academy of Sciences,\\
ul.Radzikowskiego 152, 31-342 Krak\'ow, Poland}
\author{Sebastian Kubis}
\email{skubis@pk.edu.pl}
\author{W\l odzimierz~W\'ojcik}
\affiliation{Institute of Physics, Cracow University of Technology,  Podchor\c{a}\.zych 1, 
30-084 Krak\'ow, Poland}

\begin{abstract}
The quartic term in the framework of relativistic mean field theory with inclusion of scalar meson interactions is investigated.
It is shown that the quartic term in the asymmetric expansion of nuclear matter energy may reach very large
values. This makes the even power expansion of asymmetry questionable and suggests possible non-analytic contributions to the energy of matter.  
\end{abstract}

\maketitle

\section{Introduction}

The fundamental quantity for describing neutron star matter is the energy density $\eps(n,x)$ which is a function of the baryon number $n$ and proton fraction $x=\frac{n_p}{n}$. The proton fraction is the number of protons per baryon.
The dependence of the energy of nuclear matter on the isospin asymmetry, represented by $x$,  can be described by the quantity called the {\em symmetry energy}. It plays an important role in many aspects of nuclear physics and astrophysics, such as the mass-radius relation and the composition of the crust. There are many scientific papers based on both theoretical calculations and experimental measurements, which focus on the energy symmetry behavior and its dependence on the density. Although it is relatively well determined at the nuclear saturation density $n_0=0.16 \rm ~fm^{-3}$, little is known about its behavior at densities much higher than $n_0$. 
Moreover, the asymmetry dependence of $\eps$ is also poorly known. It can be quantitatively measured  by expansion in powers of $1-2 x$.  The isospin symmetry ensures that only even powers are present:
\begin{align}
\begin{split}
\eps(n,x)=&\eps(n,1/2)+S_{2}(n)(1-2x)^2+\\
& S_{4}(n)(1-2x)^4 + \mathcal{O}((1-2x)^6).
\label{empirical-law}
\end{split}
\end{align}
As per Taylor's theorem, the coefficients of Eq.~(\ref{empirical-law}) for order $2i$ take the following forms:
\begin{equation} 
S_{2i}(n)=\left.\frac{1}{2^{2i} \cdot (2i)!}\frac{\partial^{2i} 
\left(\eps\left(n,x\right)/n \right)}{{\partial x}^{2i}}\right|_{x=\frac{1}{2}}.
\label{coef}
\end{equation}

The first term in Eq.~(\ref{empirical-law}) describes symmetric nuclear matter, which can be determined relatively well around $n_0$ using the compressibility coefficient. The rest  of this equation deals with asymmetric nuclear matter (ANM). Although details of ANM are still far from certain, great progress has occurred in recent years, particularly in the determination of the quadratic term $S_{2}$.
 The behavior of $S_2$, as a function of the density, is better understood, especially at low energy densities, due to the determination of the symmetry energy slope at $n_0$ (for a review see \cite{Horowitz:2014bja}).
The higher orders of this equation are still poorly known. There is widespread conviction that the next term in Eq.~(\ref{empirical-law}), called the {\em quartic term} is much smaller than the quadratic one and hence is usually neglected. Ignoring all terms of higher than the second order is called the {\em parabolic approximation} and is in common use. Interestingly, the quartic term may have some influence on neutron star structure and behavior, such as the density at which the core-crust transition appears or the threshold density for the URCA cooling process \cite{Steiner:2006bx,Xu:2009vi,Xu:2008vz,Seif:2013tja}.

The omission of the quartic term is usually justified by the comparison of quadratic and quartic terms
 coming from the kinetic energy contribution in the absence of any interactions
\begin{equation}
\eps_{kin} =( \int_0^{k_p}\! + \int_0^{k_n} \!) \sqrt{k^2+m^2}  \;  \frac{k^2 dk}{\pi^2} ~,
\end{equation}
then
\begin{align}
& S_2^{kin} = \frac{k^2}{6 \sqrt{k^2+m^2}} \\
& S_4^{kin} =
\frac{k^2 \left(10 k^4+11 k^2 m^2+4 m^4\right)}{648 \left(k^2+m^2\right)^{5/2}}
\end{align}
where $m$ is the  nucleon mass and $k$ is the Fermi momentum for symmetric matter $k=(\frac{3\pi^2}{2} n)^{1/3}$.
Their nonrelativistic versions take a simpler form,
\begin{align}
& S_2^{kin} = \frac{1}{3}E_{0}^{(nr)} \label{S2-nr} \\
& S_4^{kin} = \frac{1}{81} E_{0}^{(nr)} \label{S4-nr}
\end{align}
and here  $E_{0}^{(nr)}$ is the nonrelativistic Fermi energy $E_{0}^{(nr)} = \frac{k^2}{2m}$.
At the saturation point, the values of $S_2^{kin}(n_0)$ and $S_4^{kin}(n_0)$  are $12 \MeV$ and $0.5 \MeV$ respectively. However, recently presented  considerations  based on a high momentum tail in the distribution function \cite{Cai:2015mac} indicate that the kinetic energy contribution  could be much higher. 
From Eqs.~(\ref{S2-nr}) and (\ref{S4-nr}) it can be seen that they scale as $n^{2/3}$, and at very high densities, typical for a neutron star center, $S_2^{kin}$ and $S_4^{kin}$  do not exceed $50 \MeV$ and $3 \MeV$, respectively. The inclusion of the effective nucleon mass $m^*$ (which decreases with the density, and so increases these kinetic energy terms slightly), does not significantly change the situation.
Comparison of the kinetic contribution to  $S_2$ with its experimental value at saturation point $S_2(n_0)=31.2~\MeV$ indicates a large role for interactions. Moreover, the interaction contribution usually has a stronger density dependence and  becomes more significant as the density increases. 

In the relativistic mean field (RMF) framework, the quartic term has been analyzed in \cite{Lee:1998zzd}  and \cite{Cai:2011zn} and appears to be small, typically, one order of magnitude smaller than $S_2$. Other theoretical predictions for $S_4$ based on the Hartree-Fock theory 
\cite{Chen:2009wv,Xu:2010kf,Pu:2017kjx,Liu:2018far} lead to similar conclusions.

Some attempts were made to estimate the fourth order contribution to the symmetry energy at saturation density by fitting to the nuclear binding energy \cite{Jiang:2014jya,Nandi:2016xnh,Wang:2017itv} or double beta decay \cite{Wan:2018lvn}.
Results for the quartic term were unexpectedly high, as large as 20 MeV. Such uncertainty concerning the value of the quartic term gained additional complexity from recent calculations based on the chiral effective field theory. In Ref.~\cite{Kaiser:2015qia}, it was shown that the chiral expansion introduces a logarithmic asymmetry dependence $(1-2x)^4 \ln|1-2x|$ which makes the expansion given by 
Eq.~(\ref{empirical-law}) questionable.

In our work we have focused on the quartic term determination and its density dependence within the framework of the RMF  model, enriched through scalar meson interactions, e.g., by inclusion of the $\sigma\!-\!\delta$ crossing term \cite{Zabari:2019ukk}.
We show that the expansion given by  Eq.~(\ref{empirical-law}) must be treated with caution, because the quartic term  appears to be unexpectedly large. 
 
\section{model}

The Lagrangian $\mathcal{L}$ in the RMF framework includes nucleons and the four meson fields $\sigma$, $\omega$, $\rho$, and $\delta$. It is given by
\begin{equation} \label{lagrangian}
\mathcal{L} = \mathcal{L}_0 +  \mathcal{L}_{int} + \mathcal{L}_{\sigma \delta}.
\end{equation}
Additionally, we have taken into consideration the interaction between the scalar mesons $\sigma$ and $\delta$ denoted in Eq.(\ref{lagrangian})  as $L_{\sigma\delta}$. 
The first component  $\mathcal{L}_0$ includes the standard components 
for non-interacting nucleons described by a bispinor field ${\psi_p,\psi_n}$ along with mesons $\sigma$, $\omega$, $\rho$, and $\delta$. Expressions for those can be found in \cite{Zabari:2019ukk}. 
The second part of Eq.~(\ref{lagrangian}), where the coupling constants appear, is given by
\begin{equation}\label{L-int}
\mathcal{L}_{int} = g_\sigma\sigma\bar{\psi}\psi -g_\omega\omega_\mu\bar{\psi}\gamma^\mu\psi - \frac{1}{2}g_{\rho}\vec{\rho}_{\mu}\bar{\psi}\gamma^\mu\vec{\tau}\psi+g_\delta\vec{\delta}\bar{\psi}\vec{\tau}\psi -U(\sigma),
\end{equation}
where $U\left(\sigma\right)=\frac{1}{3}b m {(g_\sigma\sigma)}^3+\frac{1}{4}c {(g_\sigma\sigma)}^4$ is the self-interaction of the $\sigma$ meson.
Finally, the third component, which represents a new interaction between two scalar mesons, has the following form
 \begin{equation}\label{L-sigma-delta}
  {\cal L}_{\sigma \delta} =  \tilde{g}_{\alpha} \sigma^{\alpha} \vec{\delta}^2, ~~ \alpha =1,2.
\end{equation} 
We have considered two cases of this interaction: linear $\alpha=1$ and quadratic $\alpha=2$. Furthermore, this interaction has an influence on the field equations for the mean meson fields: $\sigma$ and 
$\delta_3$ (in the following we use the notation $\delta_3\equiv \delta$), coming from the minimum energy condition
 \begin{equation} \label{field-eq-sigma}
m_\sigma^2{\sigma}=g_\sigma\left(n_p^s+n_n^s\right)-\frac{\partial U}{\partial\sigma} -\tilde{g}_{\alpha}
\alpha\,\sigma^{\alpha-1}\delta^2,
\end{equation}
\begin{equation} \label{field-eq-delta}
m_\delta^2{{\delta}}_3=g_\delta\left(n_p^s-n_n^s\right) - 2\tilde{g}_{\alpha}\sigma^{\alpha}{\delta}.
\end{equation}
\begin{table}[t!]
\centering
\caption{Model parameters for isoscalar and isovector sectors.
\label{tab-parameters}}
\begin{tabular}{|c|l|r|r|r|r|r|}\hline 
\multirow{4}{*}{\rotatebox[origin=c]{90}{\parbox[c]{1cm}{\centering isoscalar sector}}}  & \multicolumn{6}{c|}{$C_{\sigma}^2 = 11 \rm ~ [fm^2]$}\\
~ & \multicolumn{6}{c|}{$C_{\omega}^2 = 6.48 \rm ~ [fm^2]$} \\ 
~ & \multicolumn{6}{c|}{$b = 0.054$} \\ 
~ & \multicolumn{6}{c|}{$c = - 0.0057$} \\ \hline
\multirow{20}{*}{\rotatebox[origin=c]{90}{\parbox[c]{1cm}{\centering isovector sector}}}
 & \multicolumn{6}{c|}{standard RMF theory} \\ \cline{2-7}
\\ [-2ex]
~ & \multicolumn{2}{c}{$C_{\delta}^2 \rm~[fm^2]$} & \multicolumn{2}{c}{$C_{\rho}^2 \rm~[fm^2]$} & \multicolumn{2}{c|}{$L \rm~[MeV]$}  \\ 
~ & \multicolumn{2}{c}{1.0} & \multicolumn{2}{c}{7.1} & \multicolumn{2}{c|}{91.8}  \\ 
~ & \multicolumn{2}{c}{2.5} & \multicolumn{2}{c}{12.3} & \multicolumn{2}{c|}{103.7} \\ 
~ & \multicolumn{2}{c}{3.25} & \multicolumn{2}{c}{14.8} & \multicolumn{2}{c|}{110.1}  \\ 
~ & \multicolumn{2}{c}{3.5} & \multicolumn{2}{c}{15.7} & \multicolumn{2}{c|}{112.3} \\ \cline{2-7}
~ & \multicolumn{6}{c|}{linear $\sigma$-$\delta$ meson interaction ($\alpha = 1$)}\\ 
~ & \multicolumn{6}{c|}{~~~~$g_{\alpha =1} = - 0.009 \rm ~ fm^{-1}$~~~~}\\ \cline{2-7}
\\ [-2ex]
~ & \multicolumn{2}{c}{$C_{\delta}^2 \rm~[fm^2]$} & \multicolumn{2}{c}{$C_{\rho}^2 \rm~[fm^2]$} & \multicolumn{2}{c|}{$L \rm~[MeV]$} \\ 
~ & \multicolumn{2}{c}{1.0}  & \multicolumn{2}{c}{7.4} & \multicolumn{2}{c|}{ 88.8 } \\ 
~ & \multicolumn{2}{c}{2.5}  & \multicolumn{2}{c}{14.8} & \multicolumn{2}{c|}{77.6 } \\ 
~ & \multicolumn{2}{c}{3.25}  & \multicolumn{2}{c}{19.6} & \multicolumn{2}{c|}{57.2 } \\ 
~ & \multicolumn{2}{c}{3.5}  & \multicolumn{2}{c}{21.3} & \multicolumn{2}{c|}{46.8 } \\ \cline{2-7}
~ & \multicolumn{6}{c|}{~~quadratic $\sigma$-$\delta$ meson interaction  ($\alpha = 2$)~~}\\ 
~ & \multicolumn{6}{c|}{$g_{\alpha =2} = - 0.004$} \\ \cline{2-7}
\\ [-2ex]
~ & \multicolumn{2}{c}{$C_{\delta}^2 \rm~[fm^2]$} & \multicolumn{2}{c}{$C_{\rho}^2 \rm~[fm^2]$} & \multicolumn{2}{c|}{$L \rm~[MeV]$} \\ 
~ & \multicolumn{2}{c}{1.0} & \multicolumn{2}{c}{7.3} & \multicolumn{2}{c|}{88.3} \\ 
~ & \multicolumn{2}{c}{2.5} & \multicolumn{2}{c}{13.6} & \multicolumn{2}{c|}{ 77.6 } \\ 
~ & \multicolumn{2}{c}{3.25} & \multicolumn{2}{c}{17.1} & \multicolumn{2}{c|}{ 62.3 } \\ 
~ & \multicolumn{2}{c}{3.5} & \multicolumn{2}{c}{18.4} & \multicolumn{2}{c|}{ 55.2 } \\ \hline
\end{tabular}
\end{table}

The field equations for $\omega$ and $\rho$ remain unaffected, following the standard RMF model. 
Here, the following notation has been used:
\begin{equation}
n_i = \frac{2}{{(2\pi)}^3}\int_{0}^{k_i}d^3k~~,~~~ i= p,n  
\end{equation}
\begin{equation}\label{scalar-density}
 n_i^s=\frac{2}{{(2\pi)}^3}\int_{0}^{k_i}\frac{m_i^\ast}{\sqrt{k^2+m_i^{\ast2}}}d^3k ~~,~~~ i= p,n  
\end{equation}
\begin{equation}\label{eff-mass}
 \left\{ 
\begin{array}{ll}
m_p^\ast=m-g_\sigma{\sigma}-g_\delta{{\delta}},\\
m_n^\ast=m-g_\sigma{\sigma}+g_\delta{{\delta}},
 \end{array}
  \right.
\end{equation}
where $i=p,n$ stands for proton and neutron, $n=n_p+n_n$ is the baryon density, $n_i^s$ is the scalar density, $ k_i=\left(3\ \pi^2n_i\right)^\frac{1}{3}$ denotes the Fermi momenta, and $m_i^*$ are the effective masses. Finally, the energy density has the form:
\begin{equation}\label{energy}
\begin{split}
\eps=& \frac{1}{2} m_{\sigma}^2\sigma^2+\frac{1}{2} m_{\omega}^2\omega^2+\frac{1}{2} m_{\rho}^2\rho^2+\frac{1}{2} m_{\delta}^2\delta^2+U(\sigma) \\ 
 & + g_{\alpha} \sigma^{\alpha} \delta^2 +
 \sum_{i=n,p}  \int_{0}^{k_i}k^2  \sqrt{k^2 + m_i^{\ast 2}} \frac{dk}{\pi^2}.
\end{split}
\end{equation}
All relevant parameters of the model are listed in Table \ref{tab-parameters}. They are determined by the properties of matter at the saturation point. The constants in the isoscalar sector,  $C_\sigma, C_\omega, b$, and $c$,  are chosen to obtain  the binding energy and compressibility of matter at $n_0$, $(\eps/n-m)|_{n_0} = -16$~MeV, and $K=9 n_0^2 \left.\frac{\partial(\eps/n)}{\partial n^2}\right|_{n_0}=230$~MeV, respectively. 

In the isovector sector we fit the coupling constants $C_\rho, C_\delta, g_\alpha$ to two quantities given by the experimental data: the symmetry energy $S_2(n_0)=30$~MeV and its slope $L=3 n_0 \left.\frac{\partial S_{2}}{\partial n}\right|_{n_0}$. 
For vanishing $g_\alpha$ the constants $C_\rho, C_\delta$ are correlated by the value of $S_2(n_0)=30$~MeV,
 as shown in \cite{Kubis:1997ew}, and the slope $L$ gets a value that is too large. 
Recent measurements suggest that the slope value is 60~MeV with an uncertainty $\pm 30$~MeV \cite{Oertel:2016bki}. Such low values of $L$ can be obtained in our model only by selecting a negative cross-term coupling $g_\alpha$.
 For linear and quadratic coupling we used the values $g_1 = -0.009~\rm fm^{-1} $ and $g_2 = -0.004$. The remaining two couplings $C_\rho, C_\delta$ are uniquely determined by $S_2(n_0)$ and $L$. The relations between measured quantities and coupling constants are
listed in Table \ref{tab-parameters}. A more detailed discussion of the model parameters is given in another work \cite{Zabari:2019ukk}.


\section{Quartic term}
\begin{figure}[b!]
\begin{center}
\includegraphics[width=\columnwidth]{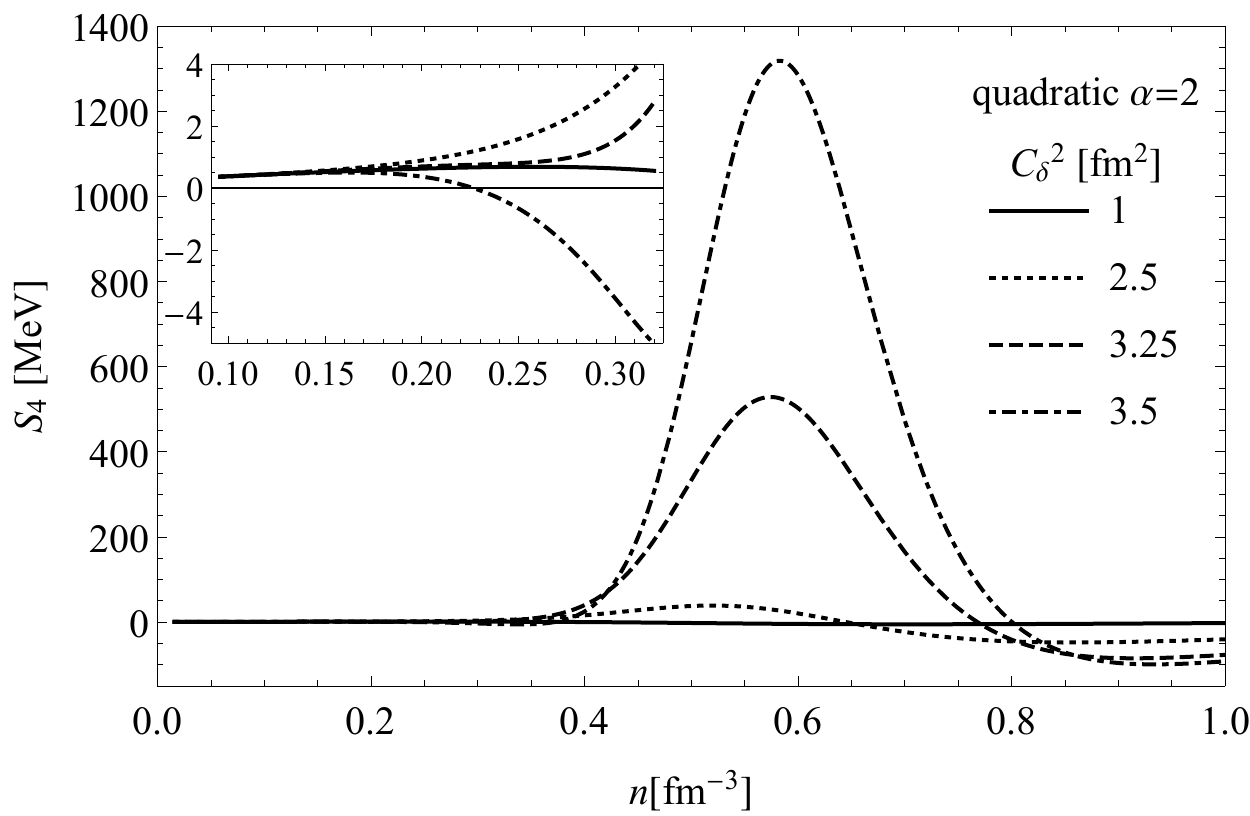}
\caption{The quartic term $S_4$ as a function of the density for changing $C_\delta$. Inset: Zoomed-in diagram of the quartic term around $n_0$.}
\label{fig:plotQzoom}
\end{center}
\end{figure}
The quartic term $S_4(n)$, which represents the fourth order of the Taylor expansion of Eq.~(\ref{empirical-law}), is defined as the fourth derivative of the energy density $\eps(n,x)$ with respect to the proton fraction $x$:
\begin{equation} \label{Q-def}
S_4(n)=\left.\frac{1}{384}\frac{\partial^4 \left(\eps\left(n,x\right)/n \right)}{{\partial x}^4}\right|_{x=\frac{1}{2}}~.
\end{equation}
The computation can be made using the energy density $\eps$ specified in Eq.~(\ref{energy}) which is characterized by meson fields $\sigma$, $\omega$, $\rho$, and $\delta$ or, alternatively, it can be expressed by the effective masses of protons and neutrons given by Eq.~(\ref{eff-mass}).
 The energy density is minimized with respect to the meson fields, which corresponds to vanishing partial derivatives  $\der{\eps}{\sigma} =0 $ and 
 $\der{\eps}{\delta}=0$ when $n$ and $x$ are kept constant, which leads to the equations of motion, (\ref{field-eq-sigma}) and (\ref{field-eq-delta}). 
 The minimum-energy condition means that the meson fields become functions of the densities 
 $\sigma(n,x), \delta(n,x)$.  The second derivative $\frac{\partial^2 \eps(n,x)}{\partial x^2}$, which corresponds to $S_{2}(n)$, holds the first derivative of the $\delta$ and $\sigma$ meson fields: $\frac{\partial \delta}{\partial x}, \frac{\partial \sigma}{\partial x}$, while the fourth derivative, which  corresponds to $S_4(n)$, includes  field derivatives up to third order.
In general, the $n$-th derivative of $\eps$ requires the $(n-1)$-th  derivative of the meson fields. 
For the isoscalar field $\sigma$, only even order derivatives remain at $x=1/2$ whereas for the isovector field $\delta$, only odd order derivatives remain. Summing up, for $S_2$ one needs to know $\frac{\partial \delta}{\partial x}$ and $S_4$ requires $\frac{\partial \delta}{\partial x}, \frac{\partial^3 \delta}{\partial x^3}, 
\frac{\partial^2\sigma}{\partial x^2}$, which can be obtained by the differentiation of the equations of motion, Eqs.~(\ref{field-eq-sigma}) and (\ref{field-eq-delta}).
In the following, we assume $k_p=k_n=k$ and $m_p^\ast=m_n^\ast=m^\ast$ for symmetric nuclear matter. Here, $k(n)={(\frac{3}{2}\pi^2n)}^{1/3}$.
The symmetry energy has the form
\begin{equation}\label{Esym-symb}
S_{2}(n)=\frac{k^2}{6 E}+\frac{1}{8}n C_{\rho}^2-C_{\delta}^2 \frac{k^3 m^{\ast 2} }{3 E^2 \pi^2 f_{\delta}},
\end{equation}
~\\
where $E=E(k,m^*)=\sqrt{k^2+m^{\ast 2}}$ and $f_\delta$ is a function depending on the nucleon-$\delta$ coupling constant and the $\sigma$-$\delta$ meson interaction coupling constant. Its form is given as
\begin{equation}\label{f-delta}
f_{\delta}= 1+C_{\delta}^2 \mathcal{A} +8 C_{\delta}^2 \sigma^{\alpha} g_{\alpha},
\end{equation}
where the $\sigma\!-\!\delta$ coupling $\tilde{g}_\alpha$, appearing in 
Eqs.~(\ref{field-eq-sigma}) and (\ref{field-eq-delta}),
 is replaced here by a combination of coupling constants
  $g_\alpha = \displaystyle\frac{\tilde{g}_\alpha}{4 g_\sigma^\alpha g_\delta^2} $,  which allows one to simplify the following formulas.
The quantity $\cal A$ is a function of scalar density $n_s=n_p^s+n_n^s$ and vector density $n=n_p+n_n$ at  $x=1/2$
\begin{equation}\label{intA}
\mathcal{A}(k, m^\ast)= \frac{4}{(2 \pi)^3} \int_{0}^{k} \frac{k^2 d^3k }{(k^2 +m^{*2})^{3/2}} 
=\frac{3 n_s}{{m^*}}-\frac{3 n}{E}~.
\end{equation}

The derivation of the quartic term is more laborious but  attainable:
\begin{widetext}
\begin{align} \label{S4}
\begin{split}
S_4 & (n)=
\frac{k^2 \left(10 k^4+11 k^2 m^{*2}+4 m^{*4}\right)}{648 E^{5}}-
g_{\alpha }^2 \frac{8 \alpha^2 m^{*4} n^3 \sigma ^{2 (\alpha -1)} C_{\delta }^8 }{E^4 f_{\delta }^4 f_{\sigma }}- \\&
 g_{\alpha} \frac{2 \alpha  m^{*2} n^2 \sigma^{\alpha -1} C_{\delta }^4 \left(6 E k^2 m^* n C_{\delta }^2 f_{\delta }+9 n C_{\delta }^4 \left(3 k^2 m^* n+2 m^{*3} n-2 E^3 n_s\right)+E^2 k^2 m^* f_{\delta }^2\right)}{3 E^7 f_{\delta }^4 f_{\sigma }} -\\&
\frac{n}{216 E^{11} k^4 f_{\delta }^4 f_{\sigma }} \left(
-3 E^5 k^8 m^{*2} f_{\delta }^4
-4 E^4 k^6 m^{*2} C_{\delta }^2 f_{\delta }^3 \left(E f_{\sigma } \left(4
   k^2+m^{*2}\right)+9 k^2 n\right) -
\right. \\&
\left.
18 E^2 k^4 m^* n C_{\delta }^4 f_{\delta }^2 \left(m^* \left(k^2 f_{\sigma } \left(k^2-2
   m^{*2}\right) \left(k^2+m^{*2}\right)+3 E n \left(3 E^4+2 k^4-4 k^2
   m^{*2}-3 m^{*4}\right)\right)-6 E^4 k^2 n_s\right) +
\right. \\&
\left.
108 E^2 k^4 m^* n^2 C_{\delta }^6 f_{\delta } \left(k^2 m^{*3} \left(E f_{\sigma }+12
   n\right)+6 E^3 k^2 n_s-9 E^4 m^* n+9 m^{*5} n\right) +
27 n^2 C_{\delta }^8 \left(
-36 E^7 k^2 n_s \left(k^2 n_s-3 E m^* n\right)+
\right. \right.  \\&
\left. \left. 
m^* \left(f_{\sigma } \left(7 k^8 m^{*3} n+2 E^7 k^4 n_s-3 E^8 k^2 m^* n\right)-9 E^4
   m^* n \left(4 k^2 m^* \left(\left(4 k^2+3 m^{*2}\right) n_s-6 E m^* n\right)+9 E^5
   n\right)\right)
\right. \right.  \\&
\left. \left. 
\left(m^*\right)^6 n \left(f_{\sigma } \left(14 k^6+10 k^4 m^{*2}+3 k^2
   m^{*4}\right)-9 E n \left(-18 E^4+16 k^4+24 k^2 m^{*2}+9
   m^{*4}\right)\right)
\right)
\right)
\end{split}
\end{align}
\end{widetext}
Its complexity is associated with the higher derivatives of the meson fields $\delta$ and $\sigma$. One may note the appearance of $f_\sigma$
\begin{equation} \label{f-sigma}
f_{\sigma}= \mathcal{A}+\frac{1}{C_{\sigma}^2}+U''(\sigma),
\end{equation}
 which includes couplings from the isoscalar sector due to the appearance of the isoscalar field derivative as explained previously.

\begin{figure}[b!]
\includegraphics[width=.8\columnwidth]{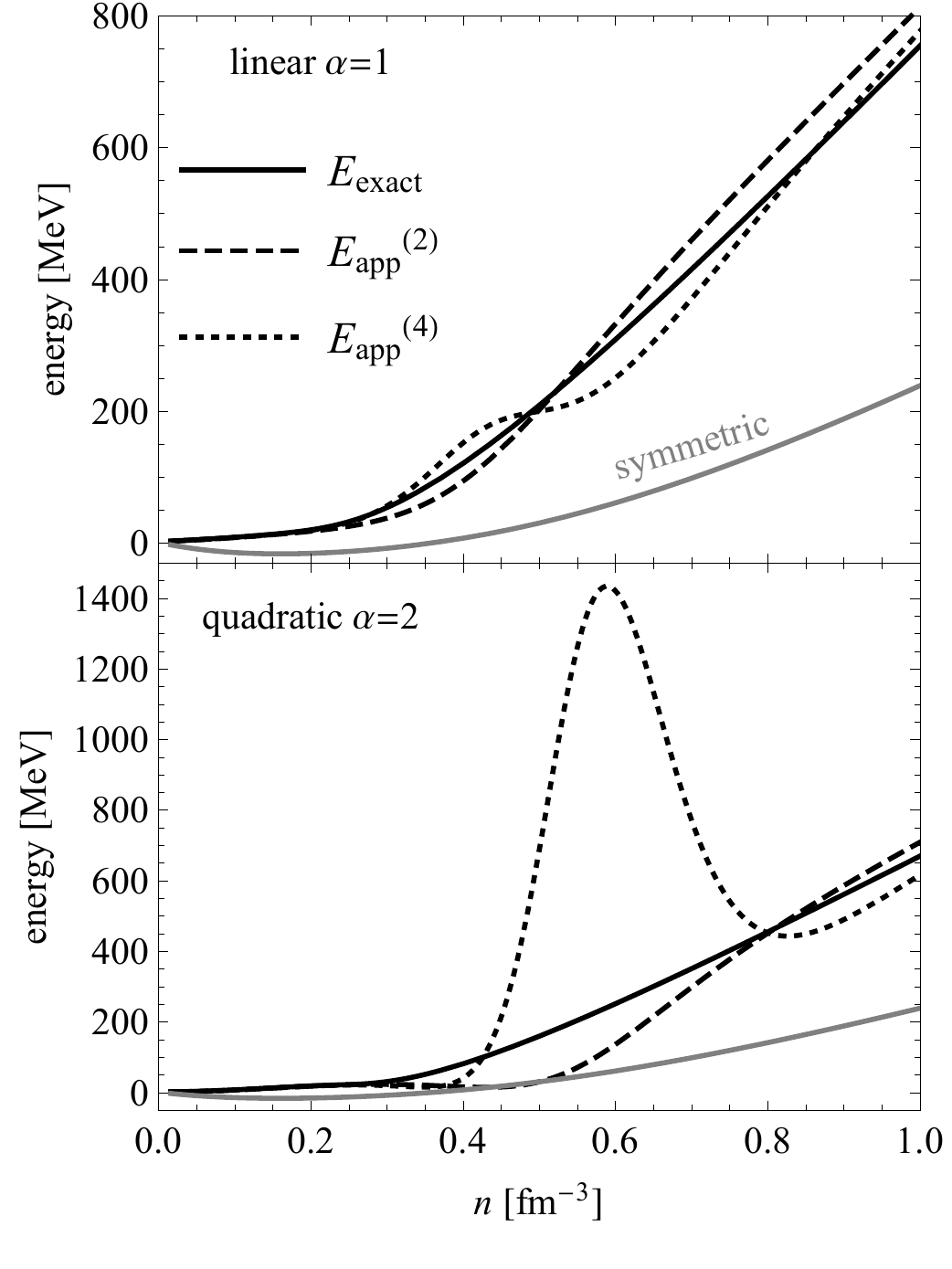}
\caption{Comparison of the approximated energy  with the exact result  for pure neutron matter for both models when $C_\delta^2 = 3.5~\rm fm^2$. The gray line represents symmetric matter.}
\label{fig:Eapp}
\end{figure}

Here we show the results for the quartic term density dependence for different $C_\delta$ coupling.
At the saturation point $S_4(n_0)$, the values remain between 0.44 and 0.65~MeV for the linear model and between 0.49 and 0.61 MeV for the quadratic model. $C_\delta$ ranges within 0 and 3.5 $\rm fm^2$. These are typical values which one may encounter in RMF models. However, as the density increases the behavior of $S_4$ changes dramatically.
As one can observe in Fig.~\ref{fig:plotQzoom}, at densities of a few times $n_0$, for the quadratic model, $S_4$ is anomalously large, reaching over a thousand MeV. In the linear model the values of the quartic term at these densities oscillate between
 -100 and 100 MeV. In order to check whether such large values are the result of erroneous derivations, we numerically calculated the fourth derivative of the energy expressed by the polynomial-interpolated function  up to sixth order. We obtained complete agreement of the numerical derivative with the analytical one given by Eq.(\ref{S4}).
  One must remember that those unexpectedly large values of the quartic term appear for the most appropriate values of $C_\delta^2=3.5 ~\rm fm^2$ which corresponds to the most likely  symmetry energy slope $L$ around 50 MeV.
When $S_4$ is much greater than the energy difference between symmetric and neutron matter, one may suspect that the expansion given by Eq.~(\ref{empirical-law})  is no longer valid. 
Indeed, in  Fig.~\ref{fig:Eapp}  the second and fourth order expansions of the energy  for pure neutron matter are compared with the exact result. For the quadratic model especially, the subsequent correction in the energy expansion is greater than the difference between symmetric and pure neutron matter. It can be seen already that the second order expansion (with parabolic approximation)
differs greatly from the exact result and adding the fourth order term does not improve it.

\begin{figure}[t]
\begin{center}
\includegraphics[width=\columnwidth]{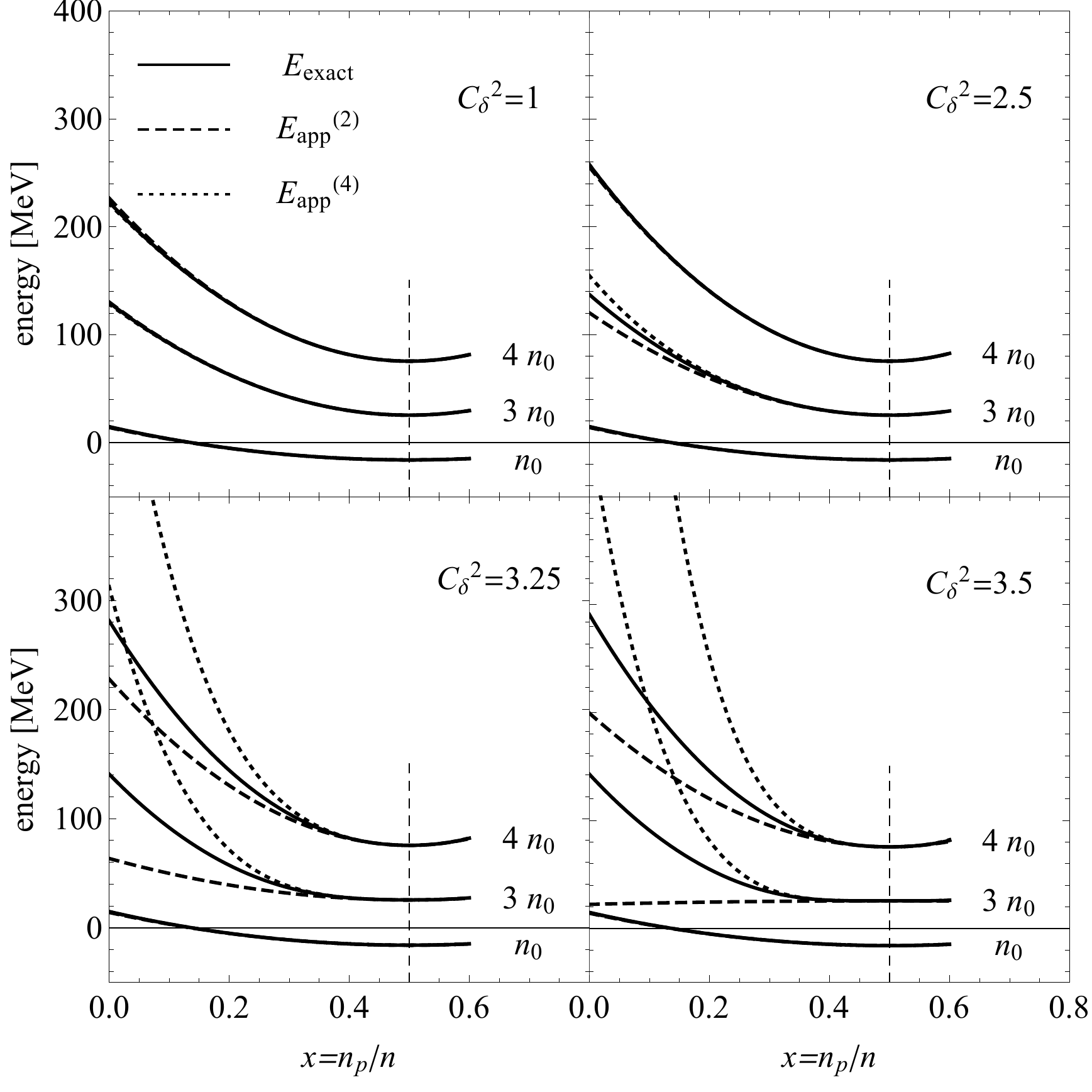}
\caption{Energy per particle in the quadratic model, $\alpha =2$, as a function of $x$ for various  densities and $C_{\delta}^2$.}
\label{fig:CrossSec}
\end{center}
\end{figure}

For further evidence we have analyzed the dependence of the energy on the proton fraction with a fixed density, shown in Fig.~\ref{fig:CrossSec}. At low densities the second and fourth order expansions work very well, they both overlap with the exact solution, independently of the coupling value $C_\delta$. However, as shown in 
Fig.~\ref{fig:plotQzoom}, at higher densities, expansion (\ref{empirical-law}) diverges significantly with increasing asymmetry.

\section{Discussion}
In this work the expansion of energy in powers of isospin asymmetry up to fourth order was studied for nuclear models with a $\sigma\smin\delta$ interaction term, Eq.(\ref{L-sigma-delta}).
 Two types of couplings were considered: linear, $\alpha=1$; and quadratic, $\alpha=2$. We focused on the density dependence of the quartic term, which appeared to be unexpectedly large at densities a bit higher than $n_0$ for both kinds of interactions. In the case of quadratic coupling the quartic term is so large that it makes the power series completely useless.  
When subsequent terms in the power expansion at a given point of a function are rapidly increasing it could be a signal that the function is not analytical at this point. Such a function can have regular derivatives only up to some
order. For example, for  $f(\beta)=\beta ^{9/2}+(1-\beta ^2)^8$ we get the expansion around 0,
$f(\beta)\approx 1 - 8 \beta^2 + 28 \beta^4$, 
where the subsequent derivatives rapidly increase. Up to fourth order the function seems to be analytical, however the fifth derivative of that function does not exist. It is difficult to say whether 
the case is the same in our model. In principle, it would be necessary to know all higher derivatives. 
Even the sixth derivative is highly complex. Nonanalyticity of the expansion of the asymmetry was recently signaled in \cite{Kaiser:2015qia}, where it was obtained from chiral perturbation theory with the inclusion of an iterated $1\pi$ exchange carried out by the contact interaction. It would be very interesting if, similarly, such nonanalyticity in the energy  could be obtained in the RMF only after the inclusion of a specific scalar meson interaction. We are aware that our final conclusion here requires further investigation. 

\bibliography{delta-sigma-qterm.bbl}

\end{document}